# Dirac charge dynamics in graphene by infrared spectroscopy

Z. Q. Li[1*], E. A. Henriksen[2], Z. Jiang[2,3], Z. Hao[4], M. C. Martin[4], P. Kim[2], H. L. Stormer[2,5,6], and D. N. Basov[1]

[1] Department of Physics, University of California, San Diego, La Jolla, California 92093, USA

[2] Department of Physics, Columbia University, New York, New York 10027, USA

[3] National High Magnetic Field Laboratory, Tallahassee, Florida 32310, USA

[4] Advanced Light Source Division, Lawrence Berkeley National Laboratory, Berkeley, California 94720, USA

[5] Department of Applied Physics and Applied Mathematics, Columbia University, New York, New York 10027, USA

[6] Bell Labs, Alcatel-Lucent, Murray Hill, New Jersey 07974, USA

*e-mail: zhiqiang@physics.ucsd.edu

**A remarkable manifestation of the quantum character of electrons in matter is offered by graphene, a single atomic layer of graphite. Unlike conventional solids where electrons are described with the Schrödinger equation, electronic excitations in graphene are governed by the Dirac Hamiltonian[1]. Some of the intriguing electronic properties of graphene, such as massless Dirac quasiparticles with linear energy-momentum dispersion, have been confirmed by recent observations[2-5]. Here we report an infrared (IR) spectromicroscopy study of charge dynamics in graphene integrated in gated devices. Our measurements verify the expected characteristics of graphene and, owing to the previously unattainable accuracy of IR experiments, also uncover significant departures of the quasiparticle dynamics from predictions made for Dirac fermions in idealized, free standing graphene. Several observations reported here indicate the relevance of many body interactions to the electromagnetic response of graphene.**

We investigated the reflectance $R(\omega)$ and transmission $T(\omega)$ of graphene samples on a $SiO_2$/Si substrate (inset of Figure 1(a)) as a function of gate voltage $V_g$ at 45K (see Methods). We start with data taken at the charge neutrality point $V_{CN}$: the gate voltage corresponding to the minimum DC conductivity and zero total charge density (inset of Fig. 1(c)). Figure 1(a) depicts $R(\omega)$ of a graphene gated structure (graphene/$SiO_2$/Si) at $V_{CN}$=3V normalized by reflectance of the substrate $R_{sub}(\omega)$. $R_{sub}(\omega)$ is dominated by a minimum around 5500 cm$^{-1}$ due to interference effects in $SiO_2$. A remarkable observation is that a monolayer of undoped graphene dramatically modifies the interference minimum of the substrate leading to a suppression of $R_{sub}(\omega)$ by as much as 15%. This observation is significant because it allows us to evaluate the conductivity of graphene near the interference structure, as will be discussed below.

Both reflectance and transmission spectra of graphene structures can be modified by a gate voltage. Figure 1 (b) and (c) display these modifications at various gate voltages normalized by data at $V_{CN}$: $R(V)/R(V_{CN})$ and $T(V)/T(V_{CN})$, where $V= V_g - V_{CN}$. These data correspond to the Fermi energy $E_F$ on the electron side and similar behavior was observed with $E_F$ on the hole side (not shown). At low voltages (<17V) we found a dip in $R(V)/R(V_{CN})$ spectra. With increasing bias this feature evolves into a peak-dip structure and systematically shifts to higher frequency. The $T(V)/T(V_{CN})$ spectra reveal a peak at all voltages, which systematically hardens with increasing bias. A voltage-induced increase in transmission ($T(V)/T(V_{CN})>1$) signals a decrease of the absorption with bias. Most interestingly, we observed that the frequencies of the main features in $R(V)/R(V_{CN})$ and $T(V)/T(V_{CN})$ all evolve approximately as $\sqrt{V}$.

In order to explore the quasiparticle dynamics under applied voltages, it is imperative to discuss first the two dimensional (2D) optical conductivity of charge neutral graphene, $\sigma_1(\omega, V_{CN})+i\sigma_2(\omega, V_{CN})$, extracted from a multilayer analysis of the devices (see Methods). Theoretical analysis[6-8] predicts a constant "universal" 2D conductivity $\sigma_1(\omega,V_{CN})=\pi e^2/2h$ for ideal undoped graphene. Our $R(\omega)/R_{sub}(\omega)$ data are consistent with this prediction. Fig. 1(a) shows a comparison between experimental $R(\omega)/R_{sub}(\omega)$ spectrum and model spectra generated assuming constant $\sigma_1(\omega,V_{CN})$ values. The constant universal conductivity offers a good agreement (within ±15%) with the experimental spectra in the range 4000-6500 cm$^{-1}$. Outside of this spectral region, our IR measurements do not allow us to unambiguously determine the absolute value of $\sigma_1(\omega, V_{CN})$; therefore the uncertainty of $\sigma_1(\omega,V_{CN})$ increases as shown by the shaded region weighted around the $\pi e^2/2h$ value. However, recent IR studies of graphene revealed a constant conductivity $\sigma_1(\omega, V_{CN}) = \pi e^2/2h$ between 2400 and 24000 cm$^{-1}$ (Ref.[9] and Mak, K.F. & Heinz, T. 2008 APS March Meeting, Abstract: L29.00006, unpublished). The universal conductivity is only weakly modified in bulk highly ordered pyrolytic graphite[10] (HOPG) and extends down to 800 cm$^{-1}$. Thus in the following discussion, we will assume $\sigma_1(\omega, V_{CN}) = \pi e^2/2h$ throughout the entire range of our data.

Electrostatic doping of graphene introduces two fundamental changes in the optical conductivity $\sigma_1(\omega,V)+i\sigma_2(\omega,V)$: a strong Drude component formed in the far-IR with $\sigma_1(\omega\rightarrow 0)=4$-$100\ \pi e^2/2h$ accompanied by a shifting of the onset of interband transitions at $2E_F$, as schematically shown in the inset of Fig. 2(b). In order to investigate these effects, we obtained $\sigma_1(\omega,V)+i\sigma_2(\omega,V)$ (Fig. 2 (b,c)) from voltage-dependent reflectance and transmission spectra (see Methods). The key features in the conductivity spectra are independent of uncertainties in $\sigma_1(\omega, V_{CN})$ discussed above. Regardless of the choice of $\sigma_1(\omega, V_{CN})$, under applied biases we observe a suppression of the conductivity compared to $\sigma_1(\omega, V_{CN})$ and a well-defined threshold structure above which the conductivity recovers the universal value $\pi e^2/2h$. The energy of the threshold structure systematically increases with voltage, a natural expectation for a transition occurring at $2E_F$. With a scattering rate $1/\tau=30$cm$^{-1}$ at 71V independently obtained from transport data, the Drude mode is rather narrow and confined below the low-ω cut-off of our measurements. We stress that the two voltage-induced transformations of the conductivity, the intraband mode and the onset of interband absorption at $2E_F$, are interdependent as suggested by our data. Indeed, assuming the intraband component can be described with a simple Drude formula $\sigma_1(\omega) = \sigma_{DC}/(1+\omega^2\tau^2)$ using $\sigma_{DC}$ and $1/\tau$ obtained from transport measurements, we find that the spectral weight removed from $\omega<2E_F$ is recovered under the Drude structure, such that the total oscillator strength given by $\int_0^{\Omega_c}\sigma_1(\omega)d\omega$ is conserved at any bias with a cutoff frequency $\Omega_c$=8000 cm$^{-1}$.

Next we extracted Fermi energy values from the $2E_F$ threshold using two different methods (see Methods). We found that the $2E_F$ values (Fig. 3(a)) are symmetric for biases delivering either holes or electrons to graphene. Moreover, $2E_F$ increases with voltage approximately as $\sqrt{V}$ (deviations from the square root law at small biases will be discussed below). Note that $E_F$ of Dirac fermions scales with the 2D carrier density N as $E_F = \hbar v_F \sqrt{\pi N}$ [2,3], where $v_F$ is the Fermi velocity. In our devices, $N = C_g V/e$ where $C_g$=115aF/μm$^2$ is the gate capacitance per unit area. Therefore, the observed $\sqrt{V}$ dependence of $2E_F$ substantiates that graphene samples integrated in gated devices are governed by Dirac quasiparticles.

Interestingly, the $2E_F$ threshold in $\sigma_1(\omega, V)$ shows a width of about 1400 cm$^{-1}$ that is independent of gate voltage and therefore of carrier density N, irrespective of a seven-fold enhancement of N between 10V and 71V. This effect is much stronger than the theoretical estimate for thermal smearing of the $2E_F$ feature at 45K[7,8], which is about 500 cm$^{-1}$. A recent theoretical study[11] showed that disorder effects and electron-phonon coupling are needed to account for the width of the $2E_F$ threshold in our data. Apart from that, a spatial variation of local $E_F$ values observed in graphene on $SiO_2$/Si substrates (Ref.[12] and Brar, V. et al., 2008 APS March Meeting, Abstract: U29.00003, unpublished) will inevitably lead to a broadening of the absorption onset at $2E_F$ in $\sigma_1(\omega)$, because IR measurements register the absorption averaged over a large area (a few microns in our experiments). The origin of the inhomogeneity of $E_F$ in graphene is still an open question[12], which needs to be explored using spatially resolved probes such as near field IR conductivity studies capable of probing the response of a material with nanometer resolution over a large area[13].

Our study has uncovered several new properties of graphene that are beyond the ideal Dirac fermion picture[14]. First, our study revealed unexpected features of $\sigma_1(\omega, V)$ below $2E_F$. The band structure of ideal graphene implies that the interband transition at $2E_F$ is the lowest electronic excitation in the system apart from the Drude response at $\omega=0$. Therefore, one anticipates finding $\sigma_1(\omega,V) \approx 0$ up to the $2E_F$ threshold, provided the Drude scattering rate is much smaller than $2E_F$. This latter condition is fulfilled for all data in Fig.2, and yet we registered significant conductivity below $2E_F$ (see supplementary information). This result has not been anticipated by theories developed for Dirac Fermions[6-8]. Both extrinsic and intrinsic effects may give rise to the residual conductivity in Fig. 2. Among the former, charged impurities and unitary scatterers (edge defects, cracks, vacancies, etc) were shown to induce considerable residual conductivity below $2E_F$[11]. However, the theoretical residual absorption in Ref. [11] is systematically suppressed with voltage, whereas this suppression was not observed in our data. In addition, the magnitude of the theoretical residual absorption is smaller compared to experimental values in Fig. 2. Therefore, it is likely that other mechanisms are also responsible for the residual conductivity in our data. One intriguing interpretation of the residual conductivity is in terms of many body interactions, which are known to produce a strong frequency dependent quasiparticle scattering rate $1/\tau(\omega)$. It is predicted theoretically that $1/\tau(\omega)$ in graphene increases with frequency due to electron-electron[15, 16] and electron-phonon interactions [11, 17]. The energy dependent scattering rate initiates a marked enhancement of the conductivity compared to the Lorentzian form prescribed by the Drude model. Such an enhancement in mid-IR frequencies has been observed in many systems[18-20].

A closer inspection of the evolution of the $2E_F$ feature with the gate voltage uncovers marked departures from the $\sqrt{V}$ dependence anticipated for Dirac quasiparticles in idealized graphene. Specifically, we observe a systematic increase of the Fermi velocity at low biases up to $v_F \sim 1.25*10^6$ m/S compared to $v_F \sim 1.10*10^6$ m/S at high doping (Fig. 3). Uncertainties in these estimates increase at low biases because in this regime the frequency of the $2E_F$ threshold is becoming comparable to its voltage independent broadening. Nevertheless, the pronounced enhancement of $v_F$ is registered in our analysis irrespective of the particular method used to extract $v_F$ from the data. This observation is indicative of a renormalization of the Fermi velocity with the enhancement of $v_F$ at low energy, which is unique for many body interactions in graphene [14, 21, 22]. Signatures of band renormalization were also observed in a previous magneto-optical study of graphene[4]. Importantly, even the smallest $v_F$ values in Fig. 3(b) are higher than that of the bulk graphite[23] (~$0.9*10^6$ m/S), which also supports the hypothesis of $v_F$ renormalization in graphene. Complimentary information on the $v_F$ renormalization in graphene can be obtained from photoemission, which is another potent probe of many body effects in solids. Currently available photoemission data were all collected for epitaxial graphene grown on SiC[24, 25]. This complicates a direct comparison with IR results for exfoliated samples on SiO$_2$/Si substrates reported here. We conclude by noting that the strong deviations of the experimental electromagnetic response from a simple single particle picture of graphene reported in our study challenge current theoretical conceptions of fundamental properties of this interesting form of carbon and also have implications for its potential applications in opto-electronics.

## Methods:

**Sample fabrication and infrared measurements**

In the graphene devices studied here, monolayer graphene mechanically cleaved from Kish graphite was deposited onto an IR transparent SiO$_2$(300nm)/Si substrate[2,3], which also serves as the gate electrode. Then standard fabrication procedures were used to define multiple Cr/Au (3/35 nm) contacts to the sample. The devices studied here exhibit mobility as high as 8700 cm$^2$ V$^{-1}$ s$^{-1}$ measured at carrier densities of ~2*10$^{12}$ cm$^{-2}$. The characteristic half-integer quantum Hall effect is observed in these samples[2,3], confirming the single layer nature of our specimen. IR experiments were carried out using an IR microscope operating with synchrotron source at the Advanced Light Source (ALS) in the frequency range of 700-8000 cm$^{-1}$. The synchrotron beam is focused in a diffraction limited spot, which is smaller than the sample. We measured the reflectance R($\omega$) and transmission T($\omega$) of the graphene devices as a function of gate voltage V$_g$ with simultaneous monitoring of the DC resistivity.

**Temperature of the graphene sample**

Data reported here were obtained in a micro-cryostat with sample mounted on a coldfinger in vacuum. The temperature of our graphene sample is warmer than that of the coldfinger, due to thermal radiation from room temperature KBr optical windows and electrical isolation of the devices from the coldfinger that compromises thermal contact. A sensor mounted in the immediate proximity to the Si substrate of the devices read T=45 K at the lowest temperatures attainable at the coldfinger. Because both the temperature sensor and the device are in nearly identical environment, we assumed this reading to be accurate for graphene as well.

**Extracting the optical constants of graphene**

The graphene device contains four layers: (1) graphene with 2D optical conductivity $\sigma(\omega) = \sigma_1(\omega) + i\sigma_2(\omega)$, (2) SiO$_2$ gate insulator, (3) Si accumulation layer that forms at the interface of SiO$_2$/Si under the applied bias and (4) Si substrate. Properties of layers 2 and 4 are independent of the gate voltage whereas layers 1 and 3 are systematically modified by Vg. In our analysis of these multilayer structures we followed the protocol detailed in Reference [26]. Specifically, we carried out reflection, transmission, and ellipsometric measurements on the Si substrates and SiO$_2$/Si wafers used in our devices and thus obtained the optical constants of layers (2) and (4). We then investigated IR properties of test devices Ti/SiO$_2$/Si as a function of gate voltage and thus extracted the optical constants of the Si accumulation layer in wafers used for graphene devices. We find that the response of the Si accumulation layer is confined to far-IR frequencies[27] and gives negligible contribution to mid-IR data in Fig.1. Finally, we employed a multi-oscillator fitting procedure[26] to account for the contribution of $\sigma(\omega)$ of graphene to the reflectance and transmission spectra shown in Fig 1 using standard methods for multilayered structures.

**Extracting Fermi energy $E_F$ from conductivity spectra**

Because of the broadening of the 2$E_F$ threshold in $\sigma_1(\omega,V)$, the $E_F$ values can be determined most accurately from the imaginary part of the optical conductivity spectra $\sigma_2(\omega,V)$ depicted in Fig.2(c). Indeed, these spectra reveal a sharp minimum at $\omega=2E_F$ in agreement with previous theoretical prediction[28]. The minimum in $\sigma_2(\omega,V)$ spectrum is found from the frequency where the derivative of $\sigma_2(\omega,V)$ with respect to frequency is zero. The uncertainties of $2E_F$ obtained from this method are related to the accuracy in defining the minimum in $\sigma_2(\omega,V)$ spectrum. Alternatively, $2E_F$ values can be extracted from the center frequency of the $2E_F$ threshold in $\sigma_1(\omega, V)$. The second method has larger uncertainties as shown in Fig. 3, due to the ambiguity of defining the center of the $2E_F$ threshold in $\sigma_1(\omega, V)$.

**Acknowledgements**

Work at UCSD is supported by DOE (No. DE-FG02-00ER45799). Research at Columbia University is supported by the DOE (No. DE-AIO2-04ER46133 and No. DE-FG02-05ER46215), NSF (No. DMR-03-52738 and No. CHE-0117752), NYSTAR, and the Keck Foundation. The Advanced Light Source is supported by the Director, Office of Science, Office of Basic Energy Sciences, of the U.S. Department of Energy under Contract No. DE-AC02-05CH11231.

**Competing Financial Interests**

The authors declare that they have no competing financial interests.

**References:**

1. Semenoff, G. W. Condensed-matter simulation of a three-dimensional anomaly. *Phys. Rev. Lett.* 53, 2449–2452 (1984)

2. Novoselov, K. S. *et al*. Two-dimensional gas of massless Dirac fermions in graphene. *Nature* 438, 197–200 (2005).

3. Zhang, Y., Tan, J. W., Stormer, H. L. & Kim, P. Experimental observation of the quantum Hall effect and Berry's phase in graphene. *Nature* 438, 201–204 (2005).

4. Jiang, Z. *et al*. Infrared spectroscopy of Landau levels of graphene. *Phys. Rev. Lett.* 98, 197403 (2007).

5. Deacon, R. S., Chuang, K.-C., Nicholas, R. J., Novoselov, K. S. & Geim, A. K. Cyclotron resonance study of the electron and hole velocity in graphene monolayers. *Phys. Rev. B* 76, 081406 (2007).

6. Ando, T., Zheng, Y., & Suzuura, H. Dynamical Conductivity and Zero-Mode Anomaly in Honeycomb Lattices. *J. Phys. Soc. Jpn.* 71, 1318-1324 (2002).

7. Peres, N. M. R., Guinea, F. & Castro Neto, A. H. Electronic properties of disordered two-dimensional carbon. *Phys. Rev. B* 73, 125411 (2006).

8. Gusynin, V. P. & Sharapov, S. G. Transport of Dirac quasiparticles in graphene: Hall and optical conductivities. *Phys. Rev. B* 73, 245411 (2006).

9. Nair, R. R. *et al*. Fine Structure Constant Defines Visual Transparency of Graphene. *Science* DOI: 10.1126/science.1156965 (2008).

10. Kuzmenko, A. B., van Heumen, E., Carbone, F. & van der Marel, D. Universal Optical Conductance of Graphite. *Phys. Rev. Lett.* 100, 117401 (2008).

11. Peres, N. M. R., Stauber, T. & Castro Neto, A.H. Conductivity of suspended and non-suspended graphene at finite gate voltage. Phys. Rev. B 78, 085418 (2008). (Preprint at <http://arxiv.org/abs/0803.2816> (2008)).

12. Martin, J. *et al.* Observation of electron–hole puddles in graphene using a scanning single-electron transistor. *Nature Phys.* 4, 144-148 (2008).

13. Qazilbash, M.M. *et al*. Mott Transition in $VO_2$ Revealed by Infrared Spectroscopy and Nano-Imaging. *Science* 318, 1750-1753 (2007).

14. Castro Neto, A. H., Guinea, F., Peres, N. M. R., Novoselov, K. S. & Geim, A. K. The electronic properties of graphene. Preprint at <http://arxiv.org/abs/0709.1163> (2007).

15. González, J., Guinea, F. & Vozmediano, M. A. H. Unconventional Quasiparticle Lifetime in Graphite. *Phys. Rev. Lett.* 77, 3589 - 3592 (1996).

16. Hwang, E. H., Hu, B. Y.-K. & Das Sarma, S. Inelastic carrier lifetime in graphene. *Phys. Rev. B* 76, 115434 (2007).

17. Park, C.-H., Giustino, F., Cohen, M.L., & Louie, S. G. Velocity Renormalization and Carrier Lifetime in Graphene from the Electron-Phonon Interaction. *Phys. Rev. Lett.* 99, 086804 (2007).

18. Basov, D.N., Singley, E.J. & Dordevic, S.V. Sum rules and electrodynamics of high-$T_c$ cuprates in the pseudogap state. *Phys. Rev. B* 65, 054516 (2002).

19. Degiorgi, L. The electrodynamic response of heavy-electron compounds. *Rev. Mod. Phys.* 71, 687-734 (1999).

20. Basov, D. N. & Timusk, T. Electrodynamics of high-$T_c$ superconductors. *Rev. Mod. Phys.* 77, 721-779 (2005).

21. Gonzalez, J., Guinea, F.. & Vozmediano, M. A. H. Marginal-Fermi-liquid behavior from two-dimensional Coulomb interaction. *Phys. Rev. B* 59, R2474- R2477 (1999).

22. Das Sarma, S., Hwang, E. H. & Tse, W.-K. Many-body interaction effects in doped and undoped graphene: Fermi liquid versus non-Fermi liquid. *Phys. Rev. B* 75, 121406 (2007).

23. Zhou, S.Y. *et al*. First direct observation of Dirac fermions in graphite. *Nature Phys.* 2, 595-599 (2006).

24. Bostwick, A., Ohta, T., Seyller, T., Horn, K. & Rotenberg, E. Quasiparticle dynamics in graphene. *Nature Phys.* **3**, 36–40 (2006).

25. Zhou, S.Y. *et al*. Substrate-induced bandgap opening in epitaxial graphene. *Nature Mat.* 6, 770-775 (2007).

26. Li, Z.Q. *et al*. Light Quasiparticles Dominate Electronic Transport in Molecular Crystal Field-Effect Transistors. *Phys. Rev. Lett.* **99**, 016403 (2007).

27. Sai, N., Li, Z.Q., Martin, M.C., Basov, D.N. & Di Ventra, M. Electronic excitations and metal-insulator transition in poly(3-hexylthiophene) organic field-effect transistors. *Phys. Rev. B* **75**, 045307 (2007).

28. Mikhailov, S. A. & Ziegler, K. New electromagnetic mode in graphene. *Phys. Rev. Lett.* 99, 016803 (2007).

## Figure Legends:

**Figure 1: The reflectance R($\omega$) and transmission T($\omega$) of a graphene device under applied gate voltages.** (a): the reflectance of the graphene device (graphene/SiO$_2$/Si) R($\omega$) normalized by that of the SiO$_2$/Si substrate R$_{sub}$($\omega$) at V$_{CN}$. A set of R($\omega$)/R$_{sub}$($\omega$) spectra generated from the multilayer model using a constant $\sigma_1(\omega, V_{CN})$ in the range of

(1±0.15)$\pi e^2/2h$ are shown as shaded area. The upper and lower boundary of the shaded area are defined by $\sigma_1(\omega, V_{CN})$ with values of 0.85*$\pi e^2/2h$ and 1.15*$\pi e^2/2h$, respectively. Inset of (a): a photograph of a graphene device together with the focused synchrotron beam (red dot). (b) and (c): $R(V)/R(V_{CN})$ and $T(V)/T(V_{CN})$ spectra of the graphene device at several voltages corresponding to $E_F$ on the electron side, where $V= V_g - V_{CN}$. Inset of (c): the smoothed DC conductivity data of the sample as a function of gate voltage Vg.

**Figure 2: The optical conductivity of graphene at different voltages.** (a), the real part of the 2D optical conductivity $\sigma_1(\omega)$ at $V_{CN}$ and 71V. The solid red line displays the region where our data support the universal result. The uncertainty of $\sigma_1(\omega, V_{CN})$ is shown by the shaded area with the theoretical $\sigma_1(\omega)=\pi e^2/2h$ plotted as dashed line. The blue dashed line is $\sigma_1(\omega)$ at 71V evaluated for the theoretical spectra: $\sigma_1(\omega, V_{CN})= \pi e^2/2h$ (red dashed line). The key spectral features of $\sigma_1(\omega, V)$ are independent of uncertainties in $\sigma_1(\omega, V_{CN})$ indicated by the shaded area, as discussed in the text. Black square on the left axis: DC conductivity at $V_{CN}$. (b) and (c), $\sigma_1(\omega)$ and $\sigma_2(\omega)$ of graphene at several voltages with respect to $V_{CN}$ corresponding to $E_F$ on the electron side based on $\sigma_1(\omega, V_{CN})=\pi e^2/2h$. The absolute values of the $\sigma_2(\omega)$ spectra in (c) have uncertainties due to the uncertainties of $\sigma_1(\omega, V_{CN})$ as discussed in the text, but the spectral features are solid. Inset of (b), the band structure of graphene near the Dirac point and the interband transition at $2E_F$.

**Figure 3: The Fermi energy $E_F$ and Fermi velocity $v_F$ of graphene.** (a), The magnitude of $2E_F$ plotted as a function of $V^{1/2}$ for the electron and hole sides with respect to the charge neutrality voltage $V_{CN}$. Red solid symbols: $2E_F$ extracted from the minimum in $\sigma_2(\omega,V)$. Blue open symbols: $2E_F$ extracted from the center of the $2E_F$ threshold in $\sigma_1(\omega, V)$. The accuracy of the obtained $2E_F$ values from the two methods is similar to the size of the symbols. Solid lines are theoretical $2E_F$ values using $v_F=1.11*10^6$ m/S. (b), $v_F$ values extracted from the $E_F$ data using theoretical formula $E_F = \hbar v_F \sqrt{\pi N}$. $v_F$ values extracted from the above two methods show similar voltage dependence.

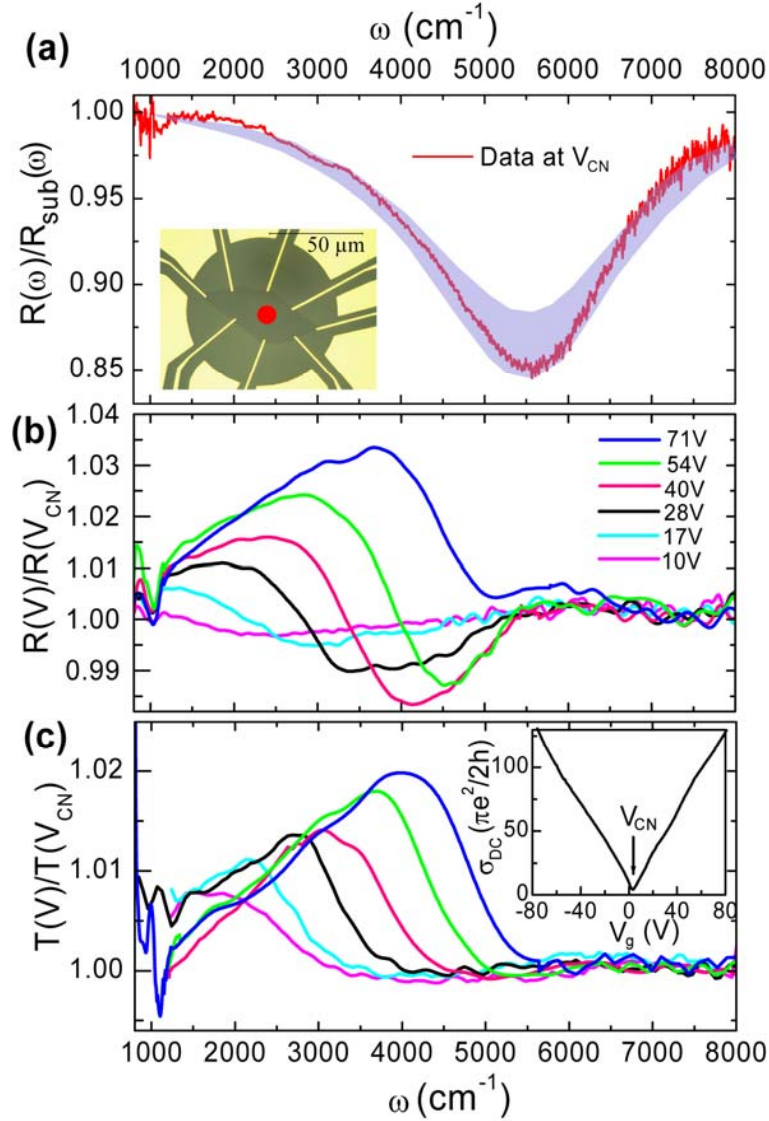

**Figure 1: The reflectance R(ω) and transmission T(ω) of a graphene device under applied gate voltages.** (a): the reflectance of the graphene device (graphene/SiO$_2$/Si) R(ω) normalized by that of the SiO$_2$/Si substrate R$_{sub}$(ω) at V$_{CN}$. A set of R(ω)/R$_{sub}$(ω) spectra generated from the multilayer model using a constant σ$_1$(ω, V$_{CN}$) in the range of (1±0.15)πe$^2$/2h are shown as shaded area. The upper and lower boundary of the shaded area are defined by σ$_1$(ω, V$_{CN}$) with values of 0.85*πe$^2$/2h and 1.15*πe$^2$/2h, respectively. Inset of (a): a photograph of a graphene device together with the focused synchrotron beam (red dot). (b) and (c): R(V)/R(V$_{CN}$) and T(V)/T(V$_{CN}$) spectra of the graphene device at several voltages corresponding to E$_F$ on the electron side, where V= V$_g$ -V$_{CN}$. Inset of (c): the smoothed DC conductivity data of the sample as a function of gate voltage Vg.

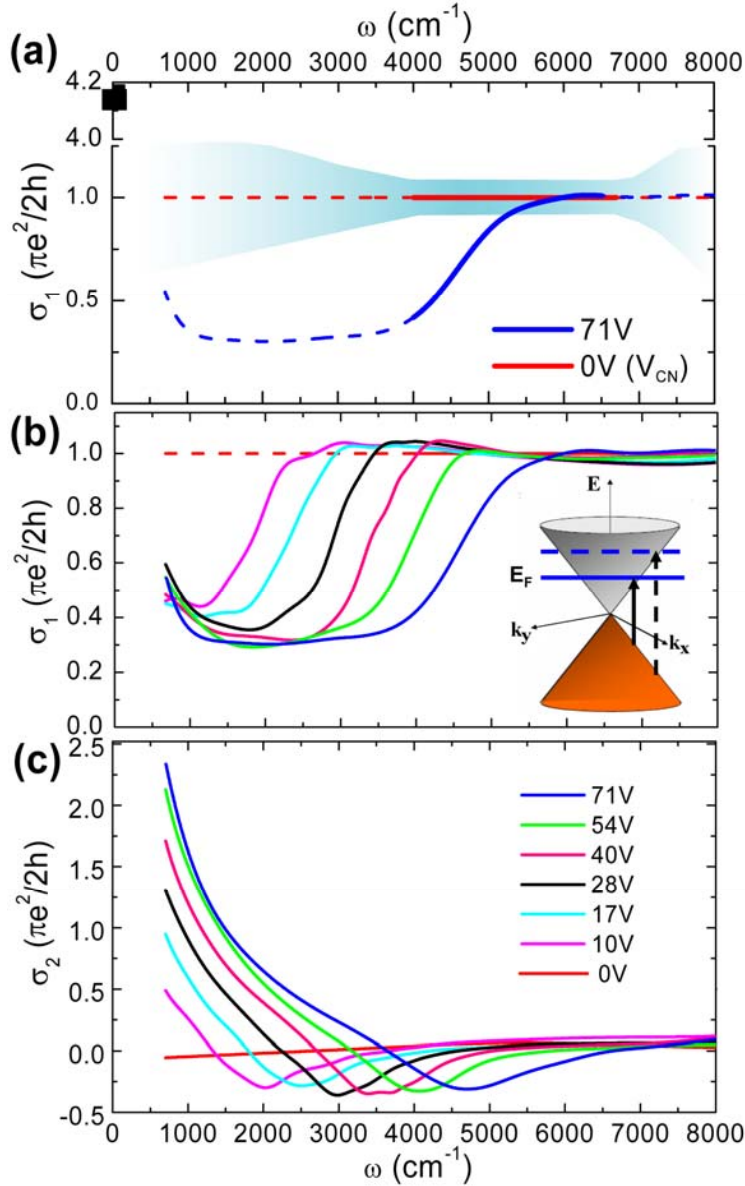

**Figure 2: The optical conductivity of graphene at different voltages.** (a), the real part of the 2D optical conductivity $\sigma_1(\omega)$ at $V_{CN}$ and 71V. The solid red line displays the region where our data support the universal result. The uncertainty of $\sigma_1(\omega, V_{CN})$ is shown by the shaded area with the theoretical $\sigma_1(\omega) = \pi e^2/2h$ plotted as dashed line. The blue dashed line is $\sigma_1(\omega)$ at 71V evaluated for the theoretical spectra: $\sigma_1(\omega, V_{CN}) = \pi e^2/2h$ (red dashed line). The key spectral features of $\sigma_1(\omega, V)$ are independent of uncertainties in $\sigma_1(\omega, V_{CN})$ indicated by the shaded area, as discussed in the text. Black square on the left axis: DC conductivity at $V_{CN}$. (b) and (c), $\sigma_1(\omega)$ and $\sigma_2(\omega)$ of graphene at several voltages with respect to $V_{CN}$ corresponding to $E_F$ on the electron side based on $\sigma_1(\omega, V_{CN}) = \pi e^2/2h$. The absolute values of the $\sigma_2(\omega)$ spectra in (c) have uncertainties due to the uncertainties of $\sigma_1(\omega, V_{CN})$ as discussed in the text, but the spectral features are solid. Inset of (b), the band structure of graphene near the Dirac point and the interband transition at $2E_F$.

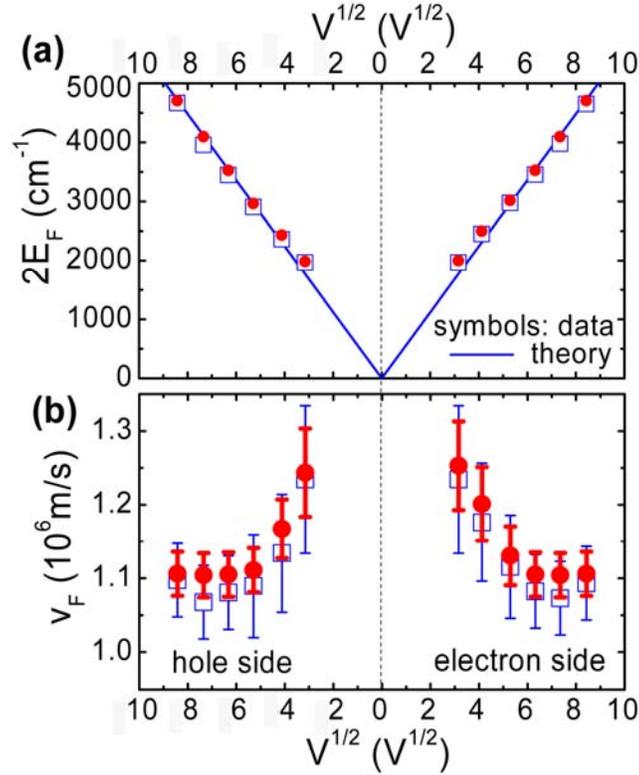

**Figure 3: The Fermi energy $E_F$ and Fermi velocity $v_F$ of graphene.** (a), The magnitude of $2E_F$ plotted as a function of $V^{1/2}$ for the electron and hole sides with respect to the charge neutrality voltage $V_{CN}$. Red solid symbols: $2E_F$ extracted from the minimum in $\sigma_2(\omega,V)$. Blue open symbols: $2E_F$ extracted from the center of the $2E_F$ threshold in $\sigma_1(\omega, V)$. The accuracy of the obtained $2E_F$ values from the two methods is similar to the size of the symbols. Solid lines are theoretical $2E_F$ values using $v_F=1.11*10^6$ m/S. (b), $v_F$ values extracted from the $E_F$ data using theoretical formula $E_F = \hbar v_F \sqrt{\pi N}$. $v_F$ values extracted from the above two methods show similar voltage dependence.

# Supplementary Information

**Spectral features in raw reflectance/ transmission data and their connection to the broadening of the 2E$_F$ threshold and residual absorption of graphene.**

As discussed in the text, our study has uncovered an anomalous width of the 2E$_F$ threshold and a strong residual absorption below 2E$_F$ in the conductivity spectra of monolayer graphene. It is straightforward to relate both effects to features in the raw data. In Fig. S1 we compare the raw R(V)/R(V$_{CN}$) and T(V)/T(V$_{CN}$) spectra (blue curves) with similar spectra generated from a model $\sigma_1(\omega)$ spectrum for ideal graphene (black curves). The top panel details the input for these model calculations. In this panel we plot with the black line the conductivity of ideal graphene $\sigma_1(\omega)$ obtained using an analytical expression for the optical constants of graphene derived by Sharapov et al.[1]:

$$\sigma_1(\Omega) = \frac{e^2 N_f}{2\pi^2 \hbar} \int_{-\infty}^{\infty} d\omega \frac{[n_F(\omega) - n_F(\omega')]}{\Omega} \frac{\pi}{4\omega\omega'} [\frac{2\Gamma(\omega)}{\Omega^2 + 4\Gamma^2(\omega)} - \frac{2\Gamma(\omega)}{(\omega+\omega')^2 + 4\Gamma^2(\omega)}](|\omega| + |\omega'|)(\omega^2 + \omega'^2)$$

where $\omega' = \omega + \Omega$, $n_F(\omega) = \frac{1}{e^{(\omega - E_F)/T} + 1}$ is the Fermi distribution, $N_f = 2$ is the spin degeneracy, and $\Gamma(\omega)$ is an impurity scattering rate. A constant scattering rate $\Gamma$ is used in the theoretical formula. In order to facilitate comparison with our mid-IR data, we have set the Fermi energy 2E$_F$=5600 cm$^{-1}$. By setting the scattering rate to $\Gamma$=1 cm$^{-1}$ and temperature to T=45 K we are able to model the threshold structure at 2E$_F$ influenced by thermal broadening representing experimental conditions. We utilized the above equation in the interband region and in order to account for the free carrier response we augmented this result with the Drude lorentzian $\sigma_1(\omega) = \sigma_{DC}/(1 + \omega^2\tau^2)$, where $\sigma_{DC}$=100* $\pi e^2/2h$ and a scattering rate 1/$\tau$=30cm$^{-1}$ is obtained from the transport data for our device. The model R(V)/R(V$_{CN}$) and T(V)/T(V$_{CN}$) spectra were calculated based on the input $\sigma_1(\omega)$ spectrum in Fig. S1(a) using the procedure described in the methods section. The dip-peak feature around 1000 cm$^{-1}$ in all the experimental and model spectra in Fig. S1(b, c) is due to a phonon of SiO$_2$. Vertical dashed lines in the plot show that the width $\delta$2E$_F$ of the interband threshold in $\sigma_1(\omega)$ is determined by broadening of the high frequency edge in T(V)/T(V$_{CN}$). Furthermore, this connection was validated through calculations using different values of the phenomenological damping constant $\Gamma$. Thus with the guidance provided by modeling results in Fig.S-1 one can read the broadening of the 2E$_F$ feature directly from the T(V)/T(V$_{CN}$) data and conclude that $\delta$2E$_F \approx$ 1400 cm$^{-1}$ at all biases.

Model spectra are equally helpful for substantiating significant residual conductivity of graphene below the 2E$_F$. For this purpose it is instructive to analyze the upper limit of the T(V)/T(V$_{CN}$) values at $\omega$=2E$_F$ corresponding to the maximum depletion of the conductivity under the applied bias. Our modeling shows that this upper limit is determined by the transmission of the graphene gated structure at the charge neutrality point T($\omega$, V$_{CN}$) and the transmission of the Si substrate T$_{sub}$($\omega$) as T($\omega$, V$_{CN}$)/T$_{sub}$($\omega$), where $T_{sub}(\omega)$ is obtained from IR measurements and $T(\omega, V_{CN})$ is calculated from the multi-layer model using the theoretical universal conductivity $\sigma_1(\omega, V_{CN}) = \pi e^2/2h$ for graphene. Provided the residual conductivity is vanishingly small, the peak in $T(V)/T(V_{CN})$ spectra at $\omega=2E_F$ reaches the upper boundary. Under these latter conditions the amplitude of peaks in a series of spectra generated for different biases will trace the boundary of the shaded region in Fig.S-1(b). However, if the depletion of the conductivity at $\omega<2E_F$ is incomplete, the residual absorption will reduce the amplitude of $T(V)/T(V_{CN})$ below the upper limit.. This is indeed the case for the experimental spectrum in Fig.S1-b taken at V=71 V and for the entire data set in Fig,1. Similarly, the amplitude of changes of reflectance is also reduced by the residual conductivity (Fig.S1-c). We note that deviations between experimental and model spectra is significant compared to the signal-to-noise of our measurements.

Here we stress that the magnitude of $\sigma_1(\omega, V)$ below the $2E_F$ threshold is sensitive to ambiguities with the choice of $\sigma_1(\omega, V_{CN})$. An assumption of the universal value for $\sigma_1(\omega, V_{CN})$ implies that the residual conductivity is as strong as $0.3*\pi e^2/2h$. Within limitations of our measurements we cannot unambiguously rule out $\sigma_1(\omega, V_{CN}) < \pi e^2/2h$ at energies below 4000 cm$^{-1}$ and dependent on the input for $\sigma_1(\omega, V_{CN})$ the residual values at $\omega<2E_F$ may significantly vary. Within these constrains, our data indicate either a breakdown of the universal conductivity $\sigma_1(\omega, V_{CN}) = \pi e^2/2h$ or significant residual conductivity below $2E_F$ at finite doping. Note that other experimental studies[2,3] attest to the validity of $\sigma_1(\omega, V_{CN}) = \pi e^2/2h$ assumption in the entire mid-IR, which implies strong residual absorption below the $2E_F$ cut-off that is nearly independent of the applied voltage.

# References:

1. Gusynin, V. P., Sharapov, S. G. & Carbotte, J. P. Unusual Microwave Response of Dirac Quasiparticles in Graphene. *Phys. Rev. Lett.* 96, 256802 (2006).

2. Mak, K.F. & Heinz, T. Private communication, unpublished.

3. Kuzmenko, A. B., van Heumen, E., Carbone, F. & van der Marel, D. Universal dynamical conductance in graphite. Preprint at <http://arxiv.org/abs/0712.0835v1> (2007).

4. Hwang, E. H., Adam, S. & Das Sarma, S. Carrier Transport in Two-Dimensional Graphene Layers. Phys. Rev. Lett. 98, 186806 (2007).

5. Tan, Y.-W. *et al.* Measurement of Scattering Rate and Minimum Conductivity in Graphene. Phys. Rev. Lett. 99, 246803 (2007).

6. Meyer, J. C. *et al*. The structure of suspended graphene sheets. *Nature* 446, 60-63 (2007).

7. Stolyarova, E. *et al*. High-Resolution Scanning Tunneling Microscopy Imaging of Mesoscopic Graphene Sheets on an Insulating Surface. *Proc. Nat. Acad. Sci. USA* 104, 9209-9212 (2007).

8. Ishigami, M., Chen, J. H., Cullen, W. G., Fuhrer, M. S. & Williams, E. D. Atomic structure of graphene on $SiO_2$. *Nano Lett.* 7, 1643–1648 (2007).

9. de Juan, F., Cortijo, A. & Vozmediano, M. A. H. Charge inhomogeneities due to smooth ripples in graphene sheets. *Phys. Rev. B* 76, 165409 (2007).

10. Guinea, F., Katsnelson, M. I. & Vozmediano, M. A. H. Midgap states and charge inhomogeneities in corrugated graphene. Preprint at <http://arxiv.org/abs/0707.0682v2> (2007).

11. Yan, J., Zhang, Y., Kim, P. & Pinczuk, A. Electric Field Effect Tuning of Electron-Phonon Coupling in Graphene. Phys. Rev. Lett. **98**, 166802 (2007).

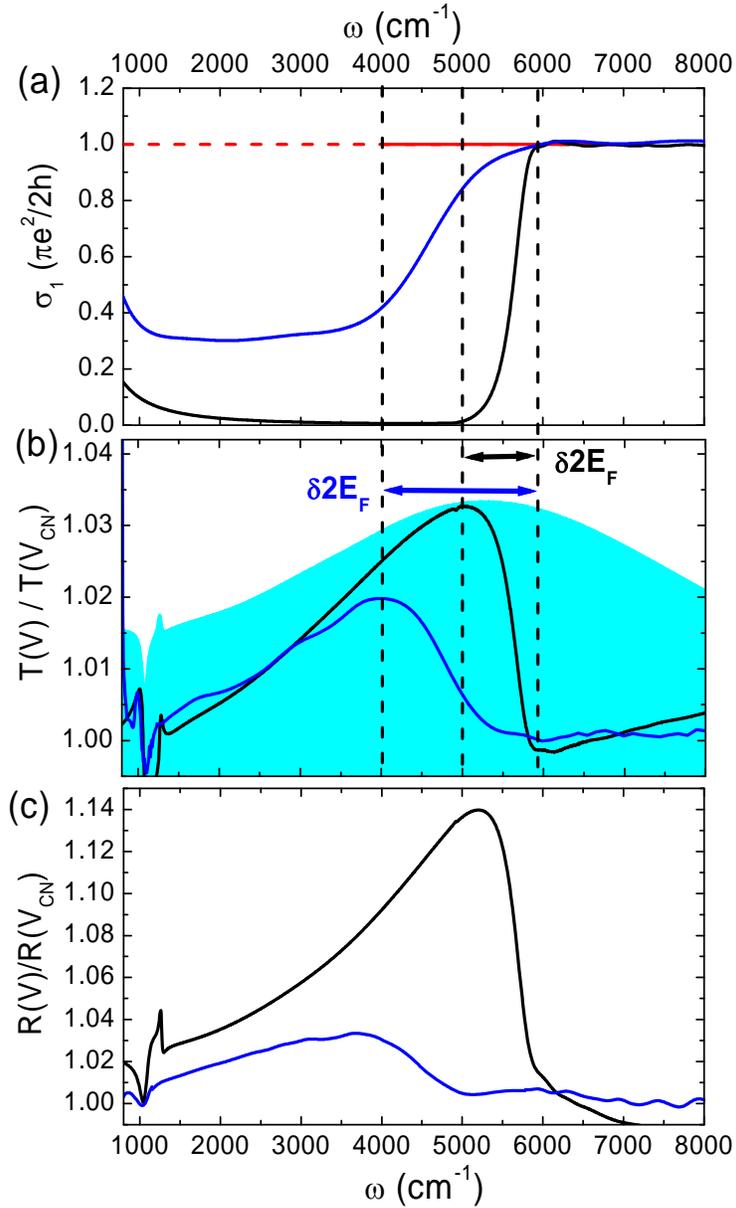

Figure S1: (a), the theoretical 2D optical conductivity $\sigma_1(\omega)$ at $V_{CN}$ (red curve) and experimental $\sigma_1(\omega)$ spectrum at 71V (blue curve), together with a model $\sigma_1(\omega)$ (black curve) with narrow width of the $2E_F$ threshold and negligible residual conductivity below $2E_F$. (b) and (c): experimental $R(V)/R(V_{CN})$ and $T(V)/T(V_{CN})$ spectra at 71V (blue spectra) and model data corresponding to the conductivity in (a) (black spectra). The upper boundary of the shaded region in (b) is the upper limit of $T(V)/T(V_{CN})$ values for different biases of our devices as described in the text. .